\documentclass[a4paper,twoside,10pt]{article}
\usepackage{graphicx,publaob}

\def\papertype{Invited Review}
\begin{document}

% Title
\title{
Supermassive binary black holes and the case of OJ 287}

% Authors
\authors{S. Komossa$^1$, S. Ciprini$^{2,3}$, L. Dey$^{4}$, L.C. Gallo$^5$, J.L. G\'omez$^{6}$, 
A. Gonzalez$^5$, } 
\authors{D. Grupe$^{7}$, A. Kraus$^1$, S.J. Laine$^{8}$, M.L. Parker$^{9,10}$, M.J. Valtonen$^{11,12}$, }
\authors{S. Chandra$^{13}$, A. Gopakumar$^{4}$, D. Haggard$^{14,15}$, \lowercase{and} M.A. Nowak$^{16}$ }  
%% \lowercase{and} et al. $^X$}

% Addresses and e-mails
% \address{$^1$Astronomical Observatory, Volgina 7, 11000 Belgrade, Yugoslavia}
\address{$^{1}$Max-Planck-Institut f\"ur Radioastronomie, Auf dem H{\"u}gel 69, 53121 Bonn, Germany       \\  %}
% \Email{skomossa}{mpifr}{de}
% \address{
$^{2}$Istituto Nazionale di Fisica Nucleare (INFN) Sezione di Roma Tor Vergata, Via della Ricerca Scientifica 1, 00133, Roma, Italy \\ 
$^{3}$ ASI Space Science Data Center (SSDC), Via del Politecnico, 00133, Roma, Italy \\
$^{4}$ Department of Astronomy and Astrophysics, Tata Institute of Fundamental Research, Mumbai 400005, India \\  
$^{5}$ Department of Astronomy and Physics, Saint Mary’s University, 923 Robie Street, Halifax, NS, B3H 3C3, Canada \\ 
$^{6}$ Instituto de Astrofísica de Andalucía-CSIC, Glorieta de la Astronomía s/n, E-18008 Granada, Spain \\  
$^{7}$Department of Physics, Earth Science, and Space System Engineering, Morehead State University, 235 Martindale Dr, Morehead, KY 40351, USA \\
$^{8}$ IPAC, Mail Code 314-6, Caltech, 1200 E. California Blvd., Pasadena, CA 91125, USA \\ 
$^{9}$ European Space Agency (ESA), European Space Astronomy Centre (ESAC), E-28691 Villanueva de la Canada, Madrid, Spain \\ 
$^{10}$ Institute of Astronomy, University of Cambridge, Madingley Road, Cambridge CB3 0HA, UK \\
$^{11}$ Finnish Centre for Astronomy with ESO, University of Turku, FI-20014, Turku, Finland \\ 
$^{12}$ Department of Physics and Astronomy, University of Turku, FI-20014, Turku, Finland \\  
$^{13}$ Centre for Space Research,
North-West University, Potchefstroom 2520, South Africa \\  
$^{14}$ Department of Physics, McGill University, 3600 rue University, Montréal, QC H3A 2T8, Canada \\  
$^{15}$ McGill Space Institute, McGill University, 3550 rue University, Montréal, QC H3A 2A7, Canada \\ 
$^{16}$ Department of Physics, Washington University in St. Louis, One Brookings Dr., St. Louis, MO 63130-4899, USA \\ 
}

% \\

% Running titles
\markboth{Binary SMBHs and the case of OJ 287}{S. KOMOSSA ET AL.}

% Abstract
\abstract{ 
Supermassive binary black holes (SMBBHs) are laboratories par excellence for relativistic 
effects, including precession effects in the Kerr metric and the emission of gravitational waves.  
Binaries form in the course of galaxy mergers, and are a key component in our understanding of galaxy evolution. Dedicated searches for SMBBHs in all stages of their evolution are 
therefore ongoing and many systems have been discovered in recent years. 
Here we provide a review of the status of observations with a focus on the multiwavelength detection methods and the underlying physics. Finally, we highlight our ongoing, dedicated multiwavelength program MOMO (for Multiwavelength Observations and Modelling of OJ 287). OJ 287 is one of the best candidates to date for hosting a sub-parsec SMBBH. The MOMO program carries out a dense monitoring at $>$13 frequencies from radio to X-rays and especially with Swift since 2015. Results so far included:  (1) The detection of two major UV-X-ray outbursts with Swift in 2016/17, and 2020; exhibiting softer-when-brighter behaviour. The non-thermal nature of the outbursts was clearly established and shown to be synchrotron radiation. (2) Swift multi-band dense coverage and XMM-Newton spectroscopy during EHT campaigns caught OJ 287 at an intermediate flux level with synchrotron and IC spectral components. (3) Discovery of a remarkable, giant soft X-ray excess with XMM and NuSTAR during the 2020 outburst. (4) Spectral evidence (at 2$\sigma$) for a relativistically shifted iron absorption line in 2020. (5) The non-thermal 2020 outburst is consistent with an after-flare predicted by the SMBBH model of OJ 287. }

% Section and subsection
\section{INTRODUCTION}

SMBBHs form in galaxy mergers which happen frequently throughout the history of the universe (Volonteri et al. 2003). 
Coalescing binaries are the
loudest sources of low-frequency gravitational waves (GWs;  Centrella et al. 2010, Kelley et al. 2019). An intense electromagnetic search for
wide and close binaries in all stages of their evolution is therefore ongoing.
%% (review by Komossa \& Zensus 2016). 
Wide pairs can be directly identified by spatially-resolved imaging spectroscopy. However, indirect
methods are required for detecting the most compact, most evolved
systems. These latter systems are well beyond the “final parsec”
in their evolution (Begelman et al. 1980; Colpi 2014; our Fig. 1) and they are in a
regime where GW emission contributes to, or even dominates, the shrinkage of their orbits. 

Detecting and modelling SMBBHs in all stages of their evolution from wide to close systems allows us to address questions which are central to our understanding of the assembly history and demography of supermassive black holes (SMBHs), and of galaxy-SMBH
formation and (co-)evolution across cosmic times (e.g., Komossa et al. 2016). These questions are:  

\begin{itemize}
    \item When does accretion start during a galaxy merger? 
    \item How long does accretion last, how much feedback is triggered,  and therefore how fast and how much do the SMBHs grow? 
     \item How much do the SMBHs’ spins change during accretion and merger?
     \item How often are both SMBHs active?
    \item How much accretion happens before and how much after the coalescence?
   \item  How efficient is the loss of angular momentum due to the binary's interactions with gas and stars (final parsec problem; Fig. 1),
   and how fast do the two SMBHs coalesce?
    \item How frequent are recoiling SMBHs and therefore, how frequent are galaxies without central SMBHs? 
    \item What is the distribution of recoil velocities and amplitudes, and how long do the SMBHs remain active after the kick? 
\end{itemize}

Further, SMBBH searches will not only inform the future space-based GW missions and the pulsar-timing arrays (PTAs) on the expected coalescence rates, but will also reveal the (pre-coalescence) initial conditions in systems that will later become major GW events. 

Whenever both merging galaxies harbor an SMBH, the formation of a binary is inevitable. The merger evolves in several stages (Begelman et al. 1980; our Fig. 1). 
The early stages of galaxy merging are driven by dynamical friction. 
At close separations, on the order of parsecs, the two SMBHs form a bound pair. The further shrinkage of their orbit then depends on the efficiency of interactions with stars and gas.
Without any such interactions which carry away energy and
angular momentum, the binary would stall and may then never coalesce within a Hubble time. This problem is known as the “final parsec problem”. 
Recent simulations have shown that interactions with gas (like molecular clouds in the center; ``massive perturbers'')  or with stars in asymmetric nuclear potentials and on elongated orbits, efficiently drive the binary beyond the final parsecs (e.g., review by Colpi 2014).   
At separations well below a parsec, emission of GWs then becomes the dominant effect that leads to  efficient further orbital
shrinkage, followed by the final coalescence. This GW-driven regime can be thought of as proceeding
in several stages: the inspiral phase, the dynamical merger, and the final ringdown. During each stage
characteristic GW radiation is emitted (Centrella et al. 2010).

% Figure (in PS or EPS format)
\begin{figure}[h]
\begin{center}
\includegraphics[width=9cm]{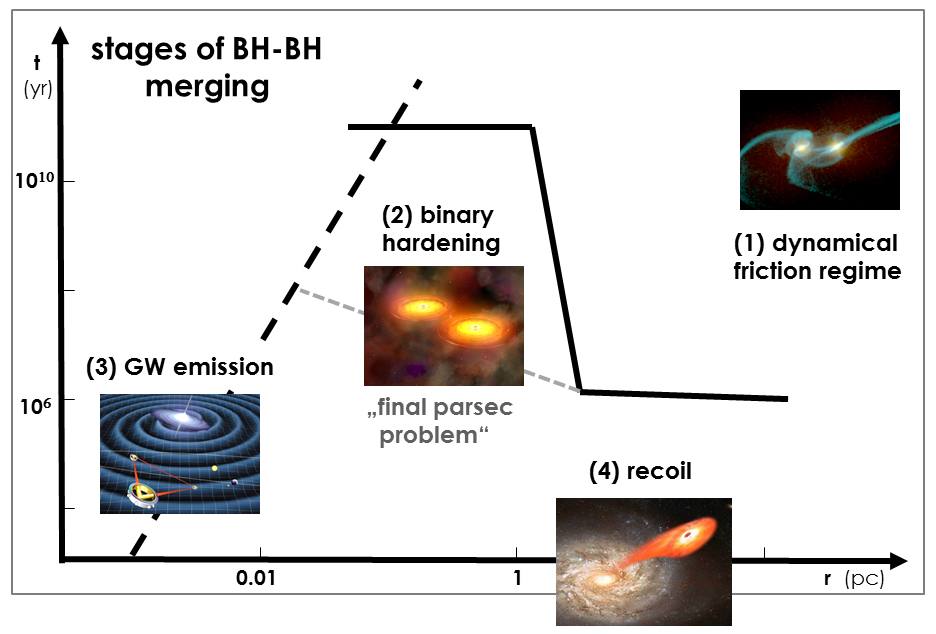}
\caption{Sketch of the evolutionary stages of SMBBHs in galaxy mergers, following Begelman et al. (1980). After (1) the merging of the galaxies due to dynamical friction, the two SMBHs will (2) form a bound pair at separations of the order of parsecs. As the binary hardens, (3) its orbit will shrink due to the emission of GW radiation, leading to the final coalescence accompanied by a strong burst of GW emission, and followed by (4) the recoil of the newly formed single black hole. The kick velocity depends on the orbital configuration and black hole mass ratio (therefore no particular timescale or distance should be associated with it in the sketch). }
\end{center}
\end{figure}

Observing pairs and binaries of SMBHs in all stages of galaxy merger evolution is of great interest.
Given the limited space of this review, and the large number of important observational and theoretical results which have emerged on this topic in recent years, it is impossible to give credit to all of these. We would therefore like to apologize in advance. We will focus on reviewing the major observational signatures which have been used to search for wide and close systems of SMBBHs, and we will mention some of the first-identified and best-studied  representative systems.

\section{Wide pairs, spatially resolved}

During gas-rich galaxy mergers, large amounts of gas are driven to the center and are available for accretion onto one or both SMBHs (Mayer 2013). In the early evolutionary stages of a galaxy merger, the two SMBHs can still be spatially resolved
(in X-rays, Chandra achieves 0.5$^{''}$  resolution, which is similar to ground-based non-AO-assisted imaging spectroscopy in the optical) and they can therefore be uniquely identified. 
Accreting SMBHs reveal their presence by a number of characteristic emission signatures. These include luminous X-ray emission from the accretion disk itself, bright and extended jets in radio(-loud) systems, and optical emission lines from the broad-line region (BLR) and narrow-line region (NLR) (these two systems of gas clouds reprocess the incident continuum emission from the accretion disk into emission lines; Peterson 1997). 

In gas-rich mergers which have their cores heavily obscured by gas and dust, X-rays are the most powerful probe of active SMBHs, since
hard X-rays can penetrate even high column densities ($N_{\rm H}$) of gas. Matter only becomes Compton-thick at $N_{\rm H} \approx 10^{24}$ cm$^{-2}$. 
Breakthroughs in the observations of galaxy pairs and mergers were made by the Chandra X-ray observatory. It was launched in 1999 and for the first time in the history of X-ray astronomy provided us with high-resolution sub-arcsecond imaging and spectroscopy. 

About 15\% of all active galactic nuclei (AGN) are radio-loud and drive powerful radio jets that are launched in the immediate vicinity of the SMBH. While jets often show a knotty structure, the true radio cores can be identified by their compactness, variability and especially their flat radio spectra.
Binaries therefore reveal themselves by the presence of two radio cores and/or two separate jet systems. 

In less heavily obscured galaxy mergers, a small fraction of photons from the accretion disk still reach the NLR, at distances of $\sim$10--1000 pc from the core. SMBH pairs therefore can reveal their presence by two NLRs in form of double-peaked emission lines and especially double-peaked [OIII]$\lambda$5007 that often is the brightest optical NLR line. Great progress in this field has been made since large spectroscopic sky surveys became available. Especially, the exceptional Sloan Digital Sky Survey (SDSS) has enabled the selection of large numbers of [OIII] double-peakers. 
At the same time, it has to be kept in mind that mechanisms other than SMBH pairs can also produce double-peaked emission lines
in {\em single} SMBH systems. These mechanisms include two-sided outflows, two-sided jet-NLR interactions, warped galactic disks, or a single active SMBH illuminating the insterstellar media of two galaxies.
%%(Xu \& Komossa 2009). 
Therefore, multiwavelength follow-up observations have been employed to confirm or reject the binary nature of [OIII] double-peakers (e.g., Fu et al. 2011, Liu et al. 2018, Comerford et al. 2018, Rubinur et al. 2019). Only a small fraction (a few percent) turned out to be active SMBH pairs.
%% Fu et al. 2012, Comerford et al. 2018 

Examples of wide systems of binary SMBHs in advanced galaxy mergers include NGC 6240 identified in X-rays with Chandra imaging spectroscopy (Komossa et al. 2003; our Fig. 2),
0402+379 (4C\,+37.11) identified in the radio regime with the VLBI technique (Rodriguez et al. 2006), 
and SDSSJ\,1502+1115 in the optical with double-peaked [OIII] emission and multiwavelength confirmation (Fu et al. 2011). 

\begin{figure}[!tbp]
%  \centering
\begin{center}
  \begin{minipage}[b]{0.43\textwidth}
    \includegraphics[width=\textwidth]{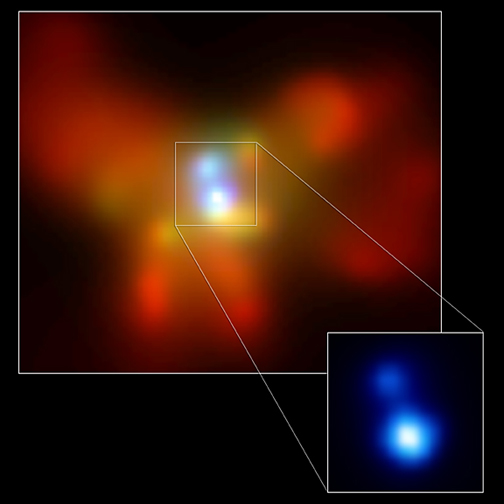}
  \end{minipage}
  \hfill
  \begin{minipage}[b]{0.51\textwidth}
    \includegraphics[width=\textwidth]{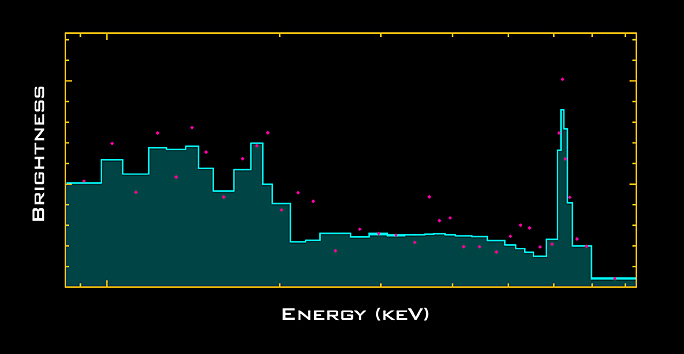}
  \end{minipage}
\end{center}
\caption{Pair of active SMBHs (blue) at the core of the galaxy NGC 6240 detected with Chandra imaging spectroscopy (image credit: NASA/CXC/Komossa et al. 2003). The two accreting SMBHs (separated by 1 kpc) are identified by their luminous, hard, point-like X-ray emission and their spectra. The Chandra ACIS-S X-ray spectra of both nuclei (right panel: Southern nucleus) show the  signatures of heavily obscured but intrinsically luminous AGN including the characteristic iron  line near 6.4 keV.   }
\end{figure}

\section{Compact binaries, spatially unresolved} 

Nearly all the sub-parsec systems are spatially unresolved, and we rely on indirect methods to identify them. The most common search methods are all based on signs of semi-periodicity and are discussed in the following sections. 

% Table
\begin{table}[h]
\footnotesize
\begin{center}
\begin{tabular}{|l|l l l l|}
\hline
name    & (AGN) type & redshift & waveband & method  \\
\hline
%\vskip
NGC 6240 &  ULIRG  & 0.024   & X-ray & imaging spectroscopy   \\ 
0402+379 & radio galaxy & 0.055 & radio & imaging spectroscopy \\
SDSS\,J1502+1115 & Seyfert &  0.39 & optical & [OIII] double-peaks \& radio imaging \\
\hline
OJ 287 & BL Lac & 0.306 & optical & semi-periodic light curve  \\
Mrk 501 & BL Lac & 0.034 & radio & semi-periodic jet structure\\
3C\,66B & radio galaxy & 0.021 & radio & semi-periodic astrometric position \\
PG\,1302--102 & FSRQ & 0.3 & optical & semi-periodic light curve\\
SDSS\,J1201+3003 & quiescent & 0.146 & X-ray & TDE lightcurve \\
NGC 4151 & Seyfert & 0.003 & optical & semi-periodic light curve \& broad line \\
\hline
\end{tabular}
\caption{Summary of the systems mentioned in this review (upper panel: spatially resolved SMBH pairs, lower panel: spatially unresolved SMBBH candidates). All of them stand out in being among the first identified and best-studied systems of their kind. Column 2 provides the classification of the host galaxy or AGN type (ULIRG stands for ultraluminous infrared galaxy; FSRQ for flat-spectrum radio quasar). In column 4, the waveband in which the system was first identified is reported. Column 5 gives the method of binary identification. }
\end{center}
\end{table}

\subsection{Semi-periodic jet structures}

A number of AGN radio jets show semi-periodic deviations from a straight line. One way to explain these observations is involving the presence
of a binary SMBH that causes either a modulation due to orbital motion of the jet-emitting SMBH around the primary SMBH, or jet precession. This method was among the first explored in the search for SMBBHs, and was motivated by the structures, bendings, and helicities observed in radio jets (Begelman et al. 1980){\footnote{See Saslaw et al. (1974) for a  discussion of SMBBH formation in three-body-interactions and radio constraints at that time.}}.
Radio interferometry has provided us with the highest-resolution observations of jets
over decades.  Mrk 501 is an early example of an AGN with helical radio jet structure, interpreted as 
Kelvin-Helmholtz instability driven at the origin through the orbital motion of an SMBBH
(Conway \& Wrobel 1995). Hydrodynamical models favored a driving period of order 10$^4$ yr to explain the observed jet morphology.

Radio observations using the technique of phase-referencing allow for ultrahigh-precision measurements of changes in the spatial location of a radio source. Such observations have the great potential of directly measuring the orbital motion of a jet-emitting SMBH (Sudou et al. 2003). 
This remarkable technique was applied to the radio galaxy 3C\,66B. Semi-periodic changes in the radio VLBA core position
at 2.3 GHz with a period of 1.05 yr were interpreted as orbital motion in a binary SMBH system (Sudou et al. 2003). Since the model requires a massive primary, part of the allowed parameter space was already excluded by pulsar timing array (PTA) constraints. These constraints place an upper limit of 
$M_{\rm BH, primary} < 1.7 \times 10^9$ M$_{\odot}$ (Arzoumanian et al. 2020).  

\subsection{Semi-periodic light curves} 

If one of the SMBHs in a binary system is emitting a (radio) jet, then there are several processes that produce a semi-periodic light-curve signal that traces either the orbital evolution of the system or else is a sign of precession induced by the second mass. 
Flux changes are then either true changes of the intrinsic emission or they are artefacts of beaming due to a jet with a systematically varying angle w.r.t our line of sight. 

An unavoidable consequence of the presence of compact SMBBHs therefore is the prediction that we should see semi-periodicities in light curves of at least a fraction of the whole (radio) binary population. However, there are also challenges when searching for periodicities in single light curves and/or large data bases: On the one hand, we need densely sampled light curves that cover at least several periods, since stochastic red noise variations can mimic periodicities (Vaughan et al. 2016). On the other hand, true intrinsic periodicities can be veiled by {\em additional} stochastic variability processes which we know are omnipresent in accretion and jet systems.

Because of the great importance of identifying sub-parsec SMBBHs, many dedicated searches are ongoing, and many candidates have been presented in the last few years. Here, we would like to highlight two examples;   
the  blazar OJ 287 (Sillanp\"a\"a et al. 1996) and the flat-spectrum radio quasar (FSRQ) PG\,1302--102 (Graham et al. 2015). OJ 287 is the best-studied and best-modelled candidate to date for hosting a compact sub-parsec binary system and will be further discussed in Sect. 4 and 5. 
PG\,1302--102 was identified in a search for periodic signals in light curves of 247000 quasars of the Catalina Transient Survey data base (Graham et al. 2015).  
The system 
triggered multiple follow-up observations and explorations of different variants of binary SMBH models (e.g., D'Orazio et al. 2015, Kovačevi{\'c} et al. 2019, Saade et al. 2020, and references therein; Fig. 3). 
Graham et al. (2015) reported a period of 5.2 yr. Their model requires jet precession in an SMBBH with $<$0.01 pc separation.   

The majority of sources with candidate semi-periodic light curves is radio-loud. 
In radio-quiet systems, 
circumbinary disk simulations of mergers predict (X-ray) emission periodically modulated at the orbital period (e.g., d'Ascoli et al. 2018). Systems with optical continuum and line variability are further discussed in Sect. 3.4. 

% Figure (in PS or EPS format)
\begin{figure}[h]
\begin{center}
\includegraphics[width=7.2cm]{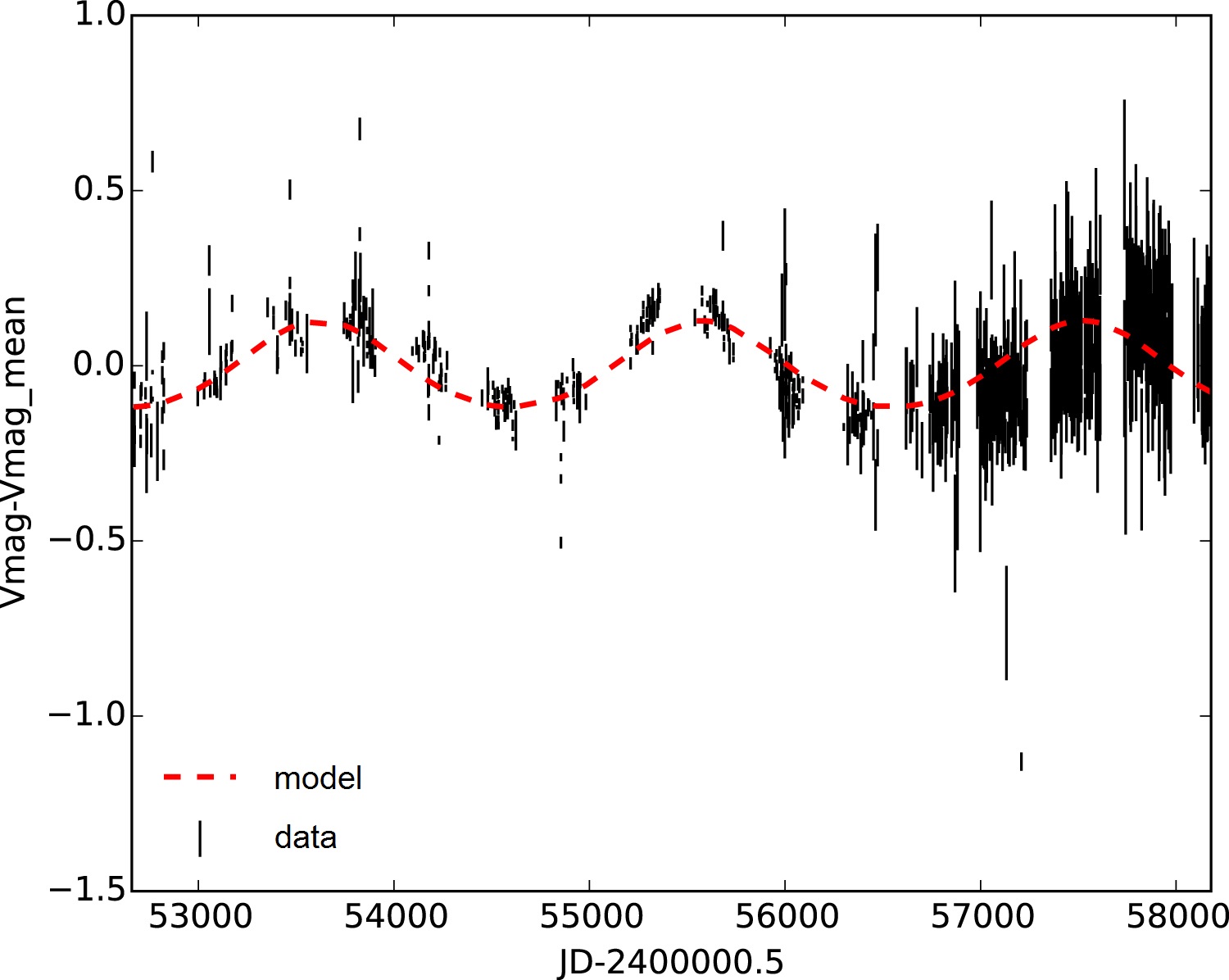}
\caption{Light curve (V magnitude, subtracted by mean value of V) of the candidate SMBBH in PG\,1302--102 and best-fit sinusoidal model, adapted from Kova{\v c}evi{\'c} et al. 2019.  
}
\end{center}
\end{figure}

\subsection{Tidal disruption event (TDE) light curves}

Stars are tidally disrupted by SMBHs once the tidal forces of the black hole exceed the self-gravity of the star. Part of the stellar material is then accreted by the SMBH causing a luminous flare of electromagnetic radiation that declines as $t^{-5/3}$ (Rees 1990). 
A few dozen TDEs have now been reported (review by Komossa 2017), following their first discovery in X-rays by the ROSAT mission.  

The TDE lightcurves of single and binary SMBHs are characteristically different. The binary model predicts characteristic dips and recoveries in a TDE light curve when
the second SMBH perturbs the stream of the stellar material, temporarily interrupting and then restarting the accretion process (Liu et al. 2009). 
This method of binary detection is of special interest, as it probes the SMBBH population in {\em quiescent}, non-active galaxies, while most other methods require at least one, or both, SMBHs to be active (an AGN) to identify the binary system. 

The first candidate SMBBH system identified from a TDE light curve is that of SDSSJ1201+3003 (Liu et al. 2014). A model with $M_{\rm BH}=10^{6-7}$ M$_{\odot}$ and mass ratio $q=0.1$ at 0.6 milli-parsec separation reproduces the light curve well.

\subsection{Double-peaked, broad Balmer lines and their variability} 

A few percent of all quasars show broad, double-peaked emission lines from the BLR. According to an early idea of Gaskell (1983), these could be the sign of the presence of a binary SMBH, with each SMBH binding its own BLR. A key prediction of the model is the Doppler-shift of each of the two line peaks as the two SMBHs orbit each other, implying a changing red/blue shift of each line peak on the timescale of years or decades. In the majority of the systems monitored, most recently selected from the SDSS spectroscopic data base, the predicted kinematic Doppler-shift was not observed, and warped disks in {\em single} AGN or other mechanisms are the preferred interpretation (Doan et al. 2020). A few candidate
binary systems have remained. Their monitoring continues. 

In a few cases, exceptional spectroscopic coverage already exists, spanning decades. 
A very well monitored AGN interpreted as binary because of its characteristic broad-line and continuum variability is NGC 4151. Based on an outstanding 43 years of spectroscopic monitoring, Bon et al. (2012)
reported periodic variability in flux and in radial velocity of one BLR component that they interpreted as shock waves generated by the supersonic motion of the components through surrounding ISM. Their model requires an SMBBH with an eccentric orbit and a period of 15.9 yr.  

\subsection{Additional methods}

Many other methods have been suggested or have already been employed to identify compact binary candidates (see also the recent review by de Rosa et al. 2019). These include: 
UV/X-Ray deficits from truncated circumbinary disks (Tanaka et al. 2012), periodic self-lensing and partial eclipses
(D'Orazio \& di Stefano 2018, Ingram et al. 2021), acceleration of jet precession (Liu \& Chen 2007),
characteristically variable and double-peaked/deformed Fe\,K$\alpha$ lines (Yu \& Lu 2001, McKernan et al. 2013), disappearance and reappearance of AGN broad Balmer lines (Wang \& Bon 2020), photocenter position variability (Popovi{\'c} et al. 2012, Kovačevi{\'c} et al. 2020),
astrometric orbital motion tracking in the Gaia data base (D'Orazio \& Loeb 2019) or with the Event Horizon Telescope (G\'omez, priv. com), magnetic field-line structure (Gold et al. 2014), (radio)-jet polarimetry (Dey et al. 2021), and  electromagnetic signals contemporaneous with binary coalescence (Haiman 2017). 
Indirectly, the detection of recoiling SMBHs also implies binary coalescences (Lousto \& Zlochower et al. 2011, Komossa et al. 2008).
In recent years, PTAs have started to place constraints on the population of binaries (Sesana et al. 2018) and were first used to constrain 3C\,66B models (Sect. 3.1).   

% Figure (in PS or EPS format)
\begin{figure}[h]
\begin{center}
\includegraphics[width=8.3cm]{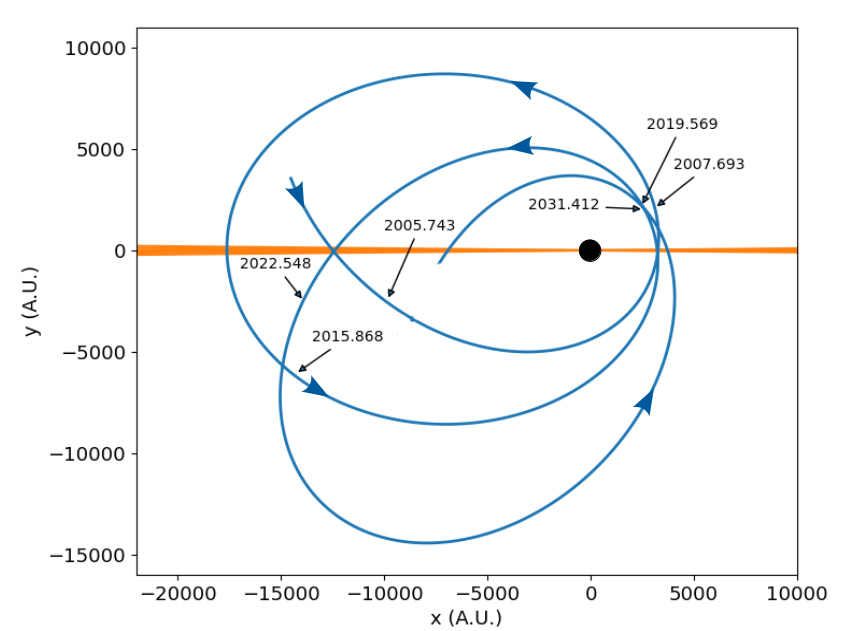}
\caption{General-relativistic, precessing orbit of the secondary SMBH in the SMBBH model of OJ 287, adopted from Dey et al. (2018) and Laine et al. (2020). 
The primary SMBH is located at the origin with its accretion disk in the y = 0 plane. 
Flares arise due to the impacts of the secondary SMBH on the accretion disk of the primary. But there is a delay (that can be calculated from the model) between the actual impacts and the times when the flares become visible.   
The black arrows point to the positions of the secondary SMBH at the times when the impact flares become visible.
}
\end{center}
\end{figure}

\section{The case of OJ 287} 

The nearby blazar OJ 287 is the longest-studied and one of the 
best candidates to date
for hosting a compact SMBBH (reviews by Kidger 2007 and Dey et al. 2019), in a regime where GW emission already contributes to a measurable shrinkage of the binary orbit (Valtonen et al. 2008, Dey et al. 2018, Laine et al. 2020). We therefore review this system in some more detail. 
The unique optical light curve of OJ 287 (e.g., Hudec et al. 2013)
shows double-peaks every $\sim$12 years that have been interpreted as arising
from the orbital motion of an SMBBH, with an orbital period on that order.
% ($\sim$9 yrs in the system's rest frame). 

Different variants of binary scenarios of OJ 287 have been considered, following the discovery that major optical outbursts of OJ 287 repeat (Sillanp\"a\"a et al. 1996). 
The best explored model by far explains the double peaks as 
episodes where the secondary SMBH impacts the disk around the primary twice during its $\sim$12 yr
orbit (``impact flares'' hereafter; Lehto \& Valtonen 1996, Valtonen et al. 2019). The most recent
4.5 order post-Newtonian orbital modelling successfully reproduces the
overall long-term light curve of OJ 287 until 2019 (Valtonen et al. 2016, Dey et al. 2018, Laine et al. 2020, and references therein).  The model requires a compact SMBBH with a semi-major axis of 9300 AU with a massive primary SMBH of $1.8\times10^{10}$
M$_{\odot}$ with spin 0.38, and a secondary of $1.5\times10^8$ M$_{\odot}$. 
Because of the strong general-relativistic orbital precession of the secondary, $\Delta \Phi$=38 deg/orbit, the impact flares are not always separated by 12 yr. Their separation varies with time and in a predictable manner (Fig. 4). 

In the SMBBH model, impact flares are triggered when the secondary SMBH crosses the accretion disk twice during its orbit. The secondary's impact drives two supersonic bubbles of hot, optically thick gas from the disk. The bubbles expand and cool. 
Once they become optically thin, they start emitting and only then the flare becomes observable
(see hydrodynamic simulations by Ivanov et al. 1998). 
Impact flares were most recently reported in 2015 and 2019 (Valtonen et al. 2016, Laine et al. 2020). At such epochs, there is an additional optical-IR emission component that may extend into soft X-rays, and the total optical flux is less polarized (Valtonen et al. 2016, Ciprini et al. 2007).   
In addition to the impact flares, the model predicts ``after-flares'' when the impact disturbance reaches the inner accretion disk (Sundelius et al. 1997) and triggers new jet activity, identified most recently with the bright X-ray--UV--optical outburst in 2020 (Komossa et al. 2020). 

% Figure (in PS or EPS format)
\begin{figure}[h]
\begin{center}
\includegraphics[clip, width=7.5cm, trim=0.7cm 2.2cm 1.7cm 0.5cm, angle=-90]{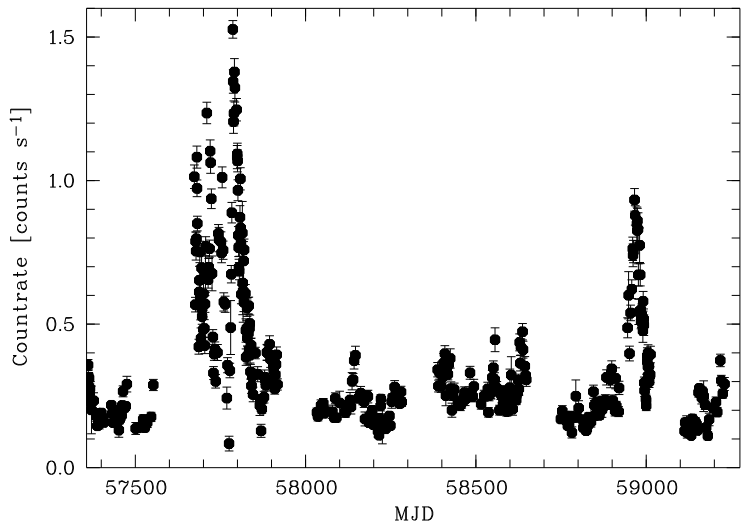}
\caption{Swift 0.3--10 keV X-ray light curve of OJ 287 since Dec. 2015, including the two bright outbursts in 2016/17 and 2020 (Komossa et al. 2017, 2020). The majority of observations was obtained in the course of the MOMO program.  }
\end{center}
\end{figure}

\section {The MOMO program}

The MOMO program, for  ``{\underline{M}}ultiwavelength {\underline{O}}bservations and {\underline{M}}odelling of {\underline{O}}J 287'' (Komossa et al. 2017, 2020, 2021) has an observational and a theoretical part. The observational part consists of long-term flux and spectroscopic monitoring and deep follow-up observations of OJ 287 at $>$13 different frequencies from the radio to the X-ray band. The Neil Gehrels Swift observatory (Swift hereafter) and the Effelsberg telescope play a central role. A few individual observations are timed with the Event Horizon Telescope (EHT; Event Horizon Telescope Collaboration et al. 2019) to obtain quasi-simultaneous SEDs{\footnote{Independent of the MOMO program, OJ 287 is a prime target of the EHT, and has been observed annually with ALMA and GMVA since 2017, providing radio VLBI observations of the twisted jet of OJ 287 at high resolution and sensitivity (G\'omez et al.  2021, in prep.).  Such observations have the  potential  of  distinguishing  between  jet  precession  triggered  by  a  binary  or a tilted  precessing accretion disk.}}.
MOMO was initiated in late 2015, with $>$2000 data sets obtained so far. The program is the densest long-term monitoring of OJ 287 involving X-rays and broad-band SEDs.  The theoretical part of the program aims at understanding aspects of accretion and jet physics of the blazar central engine in general, and the binary SMBH in particular.  Some main findings of this ongoing project are summarized below. 

% Figure (in PS or EPS format)
\begin{figure}
\begin{center}
\includegraphics[clip, trim=0.8cm 1.5cm 0.8cm 0.0cm, width=6cm,  angle=-90]{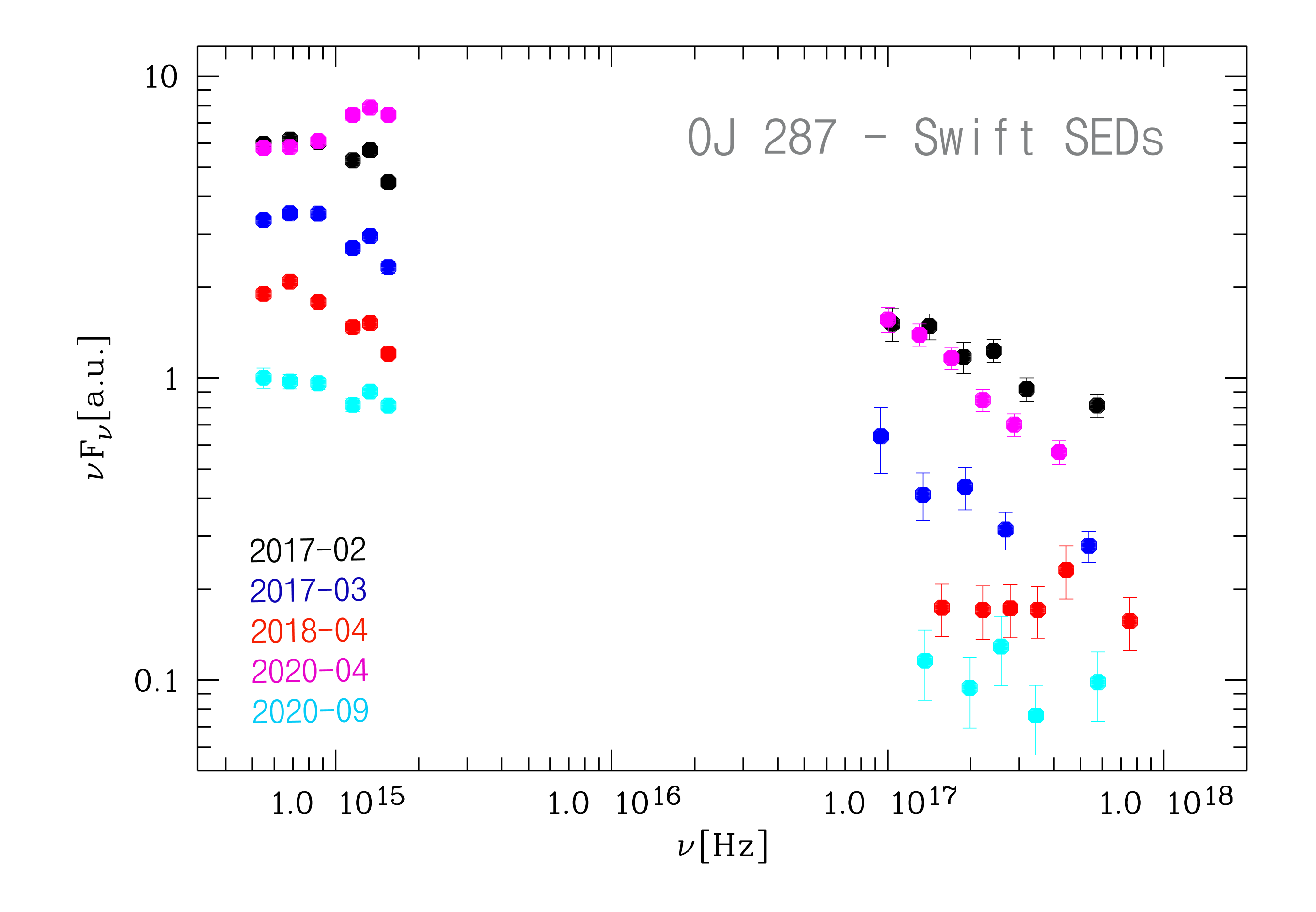}
\caption{Observed Swift SEDs of OJ 287 at selected epochs, including the 2016/17 and 2020 outbursts at peak, the March--April 2017 and 2018 near-EHT epochs, and the 2020 low-state (Komossa et al. 2021). Optical--UV and X-ray fluxes are correlated. The X-ray spectral steepening at high-states is due to the increasing contribution of the synchrotron component(s).  }
\end{center}
%%  \end{figure}

%% \begin{figure}[!tbp]
%  \centering
\begin{center}
  \begin{minipage}[b]{0.51\textwidth}
    \includegraphics[width=\textwidth]{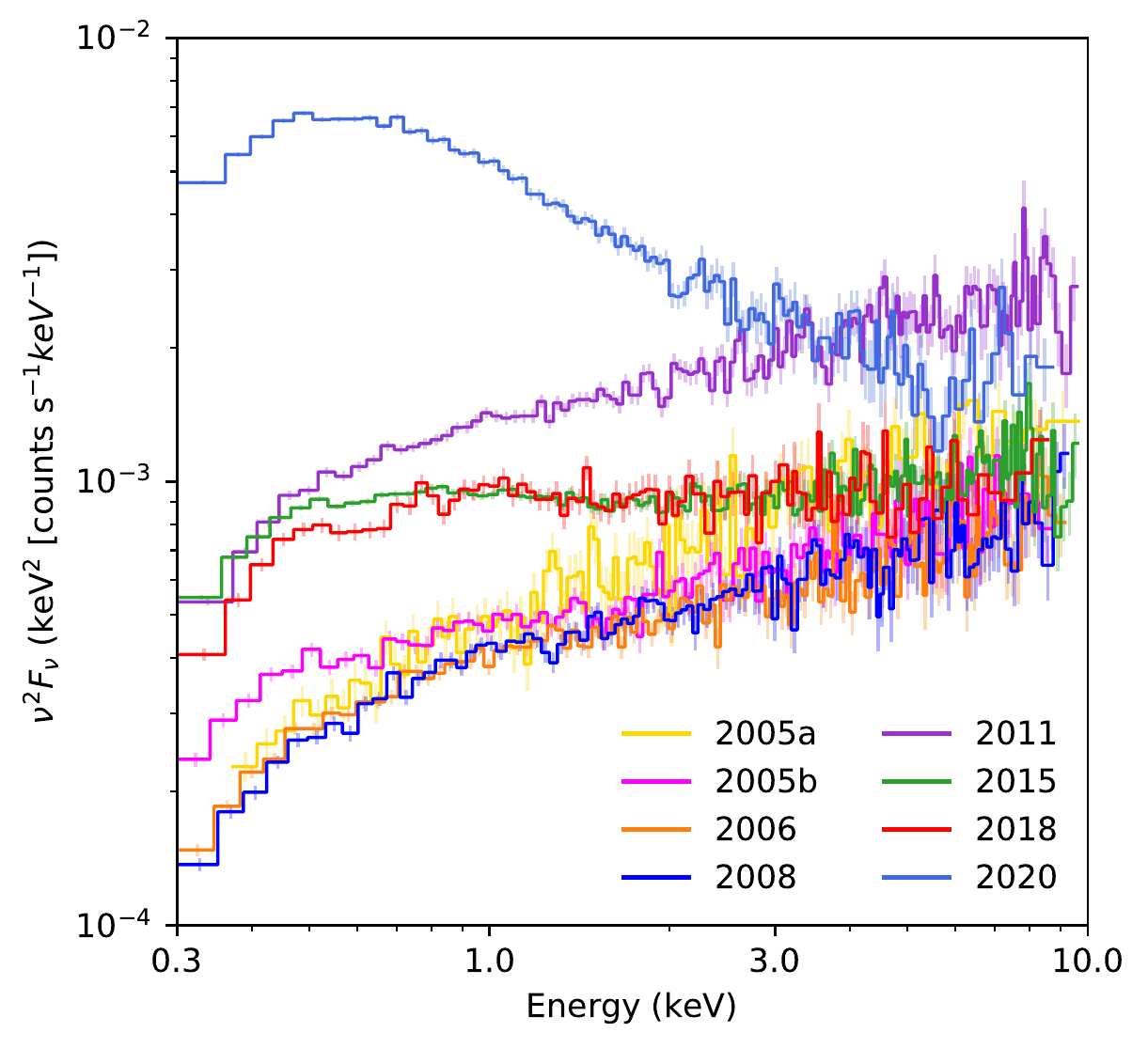}
  \end{minipage}
  \hfill
  \begin{minipage}[b]{0.48\textwidth}
    \includegraphics[width=\textwidth]{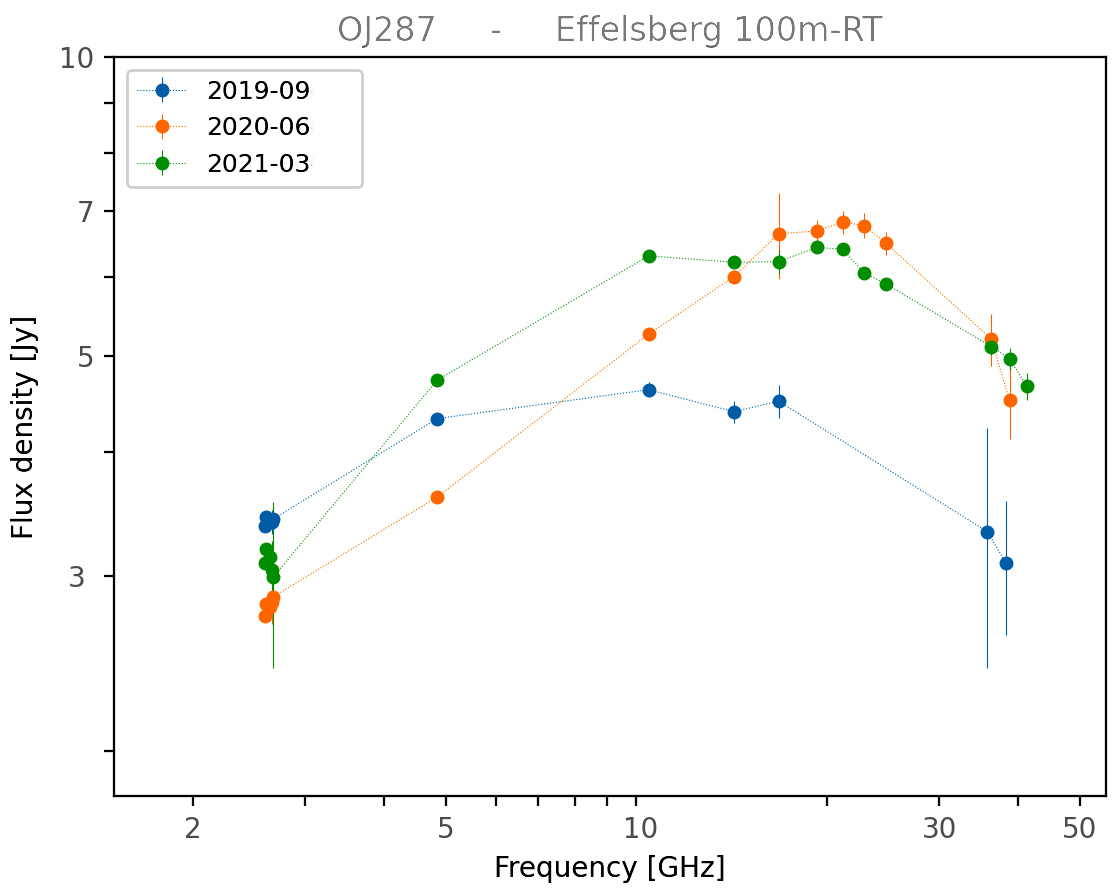}
  \end{minipage}
\end{center}
\caption{Left: Comparison of all XMM-Newton spectra of OJ 287. A giant soft X-ray excess is obvious in the 2020 spectrum (blue) obtained at the peak of the outburst
(Komossa et al. 2020). The XMM-Newton spectrum taken quasi-simultaneous with the EHT observation in 2018, at intermediate flux level, is marked in red. Right: Selected multifrequency Effelsberg radio spectra of OJ 287 between 2.6 and 40 GHz from the MOMO program. }
\end{figure}

(1) Our long-term Swift observations (Fig. 5; Komossa et al. 2017, 2020, 2021, and 2021b in prep.) established OJ 287 as one of the most spectrally variable blazars in the X-ray band (photon indices $\Gamma_{\rm x}$ = 1.5...3), changing between inverse Compton emission at low-states, and a strong synchrotron component at high-states.     

(2) Two major X-ray--UV(--optical) outbursts were discovered with Swift in 2016/17 (Komossa et al. 2017) and in 2020 (Komossa et al. 2020){\footnote{The optical outbursts were independently detected in ground-based monitoring campaigns (e.g., Zola et al. 2020)}}.  The {\em{non-thermal}} nature of the outbursts was clearly established based on multiple independent arguments: The exclusion of an accretion-disk contribution because the X-rays varied faster than the light-crossing time of the last stable orbit around the primary SMBH; the presence of a radio outburst accompanying the X-ray-optical outburst;
the close correlations of fluxes in the Swift bands; and the high level of optical polarization measured in independent projects
(Komossa et al. 2020, and references therein). 

(3) The Swift  multi-band coverage was enhanced around the dates EHT observed OJ 287 in 2017 and 2018. Selected SEDs are shown in Fig. 6. In 2018,  XMM-Newton spectroscopy (Komossa et al. 2021; our Fig. 7) during  the EHT  campaign revealed an intermediate-low flux and spectral state well described by a combination of logarithmic parabolic power-law emission of synchrotron nature and a flat ($\Gamma_{\rm x} = 1.5$) IC component.

(4)  A  remarkable,  giant  soft X-ray excess (Fig. 7) of synchrotron origin was discovered during the 2020 outburst of OJ 287 based on XMM-Newton and NuSTAR observations (Komossa et al. 2020){\footnote{A similar soft emission component was detected with Swift during the 2016/17 outburst (Komossa et al. 2017, 2020), but we lack deeper XMM-Newton observations at that epoch.}}.  
NuSTAR also revealed an additional and unusually soft emission component extending up to $\sim$ 70 keV of unknown nature{\footnote{ 
A mix of synchrotron and IC emission is a possibility; another one is a temporary accretion-disk corona contribution (even though there is no other optical--X-ray evidence for significant disk emission during the outbursts).}}. % 
Spectral evidence (at 2$\sigma$)  for  a  relativistically  shifted  iron  absorption  line  in  2020 was seen with XMM-Newton, however it needs independent confirmation in deeper future observations that catch OJ 287 in the same state.   
The 2020 X-ray--optical outburst was accompanied by a radio outburst (Fig. 7, right panel).  

(5) The non-thermal 2020 outburst is consistent with an after-flare predicted by the SMBBH model, where new jet activity is launched following a change in the accretion rate as a consequence of the secondary's disk impact (Komossa et al. 2020). 

The MOMO program will continue observations of OJ 287 as it nears its next impact flare predicted by the SMBBH model (Dey et al. 2018), expected in 2022.  

\section{Summary and outlook} 

Binary SMBHs in all stages of their evolution are central to 
SMBH demographics and galaxy evolution across cosmic times. The field has rapidly evolved in the last decade, with many systems and candidates identified through multiwavelength observations and orbital modelling, including candidate evolved systems well beyond the final parsec and wide pairs in the early stages of galaxy mergers. The fact that many of these systems are in nearby galaxies (Tab. 1) implies that binaries should be common throughout the universe. Binary SMBHs are most easily detectable electromagnetically, if at least one or both SMBHs are active. However, an elusive population of binaries could well exist at the cores of quiescent, inactive galaxies. Stellar tidal disruption event lightcurves provide us with a unique tool of searching for such a binary population.  While the last few years have seen the first direct detection of gravitational waves from stellar-mass black-hole binaries with {\em ground-based} detectors, supermassive black-hole binaries are the loudest known sources of GWs detectable with the future {\em space-based} gravitational-wave interferometer LISA.
Meanwhile, pulsar-timing arrays have greatly advanced in recent years and have started to place constraints on the population of the most massive binaries known.
The EHT with its unprecedented spatial resolution holds the promise of spatially resolving the small-separation SMBBH of OJ 287 for the first time.  
Among the population of sub-parsec binary SMBHs, the blazar OJ 287 stands out as the longest-studied and best-studied candidate which is already in a regime where gravitational-wave emission contributes measurably to the orbital shrinkage. 
As a bright multimessenger source, OJ 287 is the target of an ongoing, dense monitoring program, MOMO. The program has revealed high-amplitude outbursts interpreted in the context of the binary model as after-flares. MOMO observations continue as OJ 287 nears its next predicted impact flare. 

\vskip 0.5cm 

\noindent {\sl{Acknowledgements.}}
SK would like to thank the organizers for their invitation to give this review at the 
COST workshop on ``The gravitational-wave Universe'' during the excellent XIX Serbian Astronomical Conference. We would like to thank A.B. Kova{\v c}evi{\'c} for her permission to show her figure. This review is based on published results obtained by the international community with a multitude of observing facilities and surveys world-wide. Previously unpublished data shown here for the first time were obtained with the Neil Gehrels Swift observatory and the 100-m telescope of the Max-Planck-Institut
f\"ur Radioastronomie at Effelsberg.  
%\end{acknowledgements}

%% Equation
%\begin{equation}
%G(x)=\int_0^\infty e^{-{{x^2}\over{2}}} dx
% \end{equation}

% \vfil\newpage

% References
\references
%Hewitt, A., Burbrdge, G. : 1989, \journal{Astrophys. J. Suppl. Series}, \vol{75}, 297.

Arzoumanian, Z., et al.: 2020, \journal{Astrophys. J.} \vol{900}, 102 

Begelman, M.C., Blandford, R.D., Rees, M.J.: 1980, \journal{Nature} \vol{287}, 307

Bon, E., et al.: 2012, \journal{Astrophys. J.} \vol{759}, 118

Centrella, J., 
et al.: 2010, \journal{Reviews of Modern Physics} \vol{82}, 3069

Ciprini, S., et al.: 2007, \journal{MmSAI} \vol{78}, 741

Colpi, M.: 2014, \journal{Space Sci. Rev.} \vol{183}, 189

Comerford, J., et al.: 2018, \journal{Astrophys. J.} \vol{867}, 66

Conway, J. E., Wrobel, J. M.: 1995, \journal{Astrophys. J.} \vol{439}, 98

d'Ascoli, S., et al.: 2018, \journal{Astrophys. J.} \vol{865}, 140

de Rosa, A., et al.: 2019, \journal{New Astron. Rev.} \vol{86}, 101525

Dey, L., et al.: 2018, \journal{Astrophys. J.} \vol{866}, 11

Dey L., et al.: 2019, \journal{Universe} \vol{5}, 108

Dey, L., et al.: 2021, \journal{MNRAS} \vol{503}, 4400

Doan, A., et al.: 2020, \journal{MNRAS} \vol{491}, 1104 
D'Orazio, D.J., Haiman, Z., Schiminovich, D.: 2015, \journal{Nature} \vol{525}, 351

D’Orazio, D.J., Di Stefano, R.: 2018, \journal{MNRAS} \vol{474}, 2975 

D'Orazio, D.J., Loeb, A.: 2019, \journal{Phys. Rev. D} \vol{100}, 103016

Event Horizon Telescope Collaboration et al.: 2019, \journal{Astrophys. J.} \vol{875}, L1

Fu, H., et al.: 2011, \journal{Astrophys. J.} \vol{740}, L44

Gaskell, M.: 1983, in \journal{Proceedings of the 24th Liege Int. Astrophys. Coll.}, 473

Gold, R., et al.: 2014, \journal{Phys. Rev. D} \vol{89}, 064060

Graham, M. J., et al.: 2015, \journal{Nature} \vol{518}, 74

Haiman, Z.: 2017, \journal{Phys. Rev. D} \vol{96}, 023004

Hudec, R., et al.: 2013, \journal{Astron. Astrophys.} \vol{559}, 20

Ingram, A., et al.: 2021, \journal{MNRAS} \vol{503}, 1703

Ivanov, P.B., Igumenshchev I.V., Novikov I.D.: 1998, \journal{Astrophys. J.} \vol{507}, 131

Kelley, L., et al.: 2019, \journal{BAAS} \vol{51}, 490

Kidger, M.: 2007, \journal{Cosmological Enigmas: Pulsars, Quasars, and Other Deep-Space Questions}, The Johns Hopkins University Press

Komossa, S.: 2017, \journal{Astron. Nachrichten} \vol{338}, 256

Komossa, S., et al.: 2003, \journal{Astrophys. J. Letters} \vol{582}, L15

Komossa, S., Zhou, H., Lu, H.: 2008, \journal{Astrophys. J. Letters} \vol{678}, L81

Komossa, S., Baker, J.G., Liu, F.K.: 2016, \journal{IAU Focus Meeting} \vol{29B}, 292 

Komossa, S., et al.: 2017, \journal{IAUS} \vol{324}, 168 

Komossa, S., et al.: 2020, \journal{MNRAS} \vol{498}, L35 

Komossa, S., et al.: 2021, \journal{MNRAS}, in press 

Kova{\v c}evi{\'c}, A.B.,  et al.: 2019, \journal{Astrophys. J.} \vol{871}, 32

Kova{\v c}evi{\'c}, A.B., et al.: 2020, \journal{Astron.  Astrophys.} \vol{644}, 88

Laine, S., et al.: 2020, \journal{Astrophys. J. Letters} \vol{894}, L1 

Lehto, H.J., Valtonen, M.J.: 1996, \journal{Astrophysical Journal} \vol{460}, 207

Liu, F.K., Chen, X.: 2007, \journal{Astrophys. J.} \vol{671}, 1272

Liu, F.K., Li, S., Chen, X.: 2009, \journal{Astrophys. J. Letters} \vol{706}, L133

Liu, F.K., Li, S., Komossa, S.: 2014, \journal{Astrophys. J.} \vol{786}, 103

Liu, X., et al.: 2018, \journal{Astrophys. J.} \vol{854}, 169

Lousto, C., Zlochower, Y.: 2011, \journal{Phys. Rev. Lett.} \vol{107}, 231102 

Mayer, L.: 2013, \journal{CQGra} \vol{30}, 244008

McKernan, B., et al.: 2013, \journal{MNRAS} \vol{432}, 1468

Peterson, B.M.: 1997, \journal{An Introduction to Active Galactic Nuclei}, Cambridge University Press 

Popovi{\'c}, L.{\v C}., et al.: 2012, \journal{Astron.  Astrophys.} \vol{538}, 107

Rees, M. J.: 1990, \journal{Science} \vol{247}, 817

Rodriguez, C., et al.: 2006, \journal{Astrophys. J.} \vol{646}, 49

Rubinur, K., Das, M., Kharb, P.: 2019, \journal{MNRAS} \vol{484}, 4933

Saade, M.L., et al.: 2020, \journal{Astrophys. J.} \vol{900}, 148

Saslaw, W.C., Valtonen, M. J., Aarseth, S.J.: 1974, \journal{Astrophys. J.} \vol{190}, 253

Sesana, A., et al.: 2018, \journal{Astrophys. J.} \vol{856}, 42

Sillanp\"a\"a, A., et al.: 1996, \journal{Astron. Astrophys.} \vol{305}, L17

Sudou, H., et al.: 2003, \journal{Science} \vol{300}, 1263

Sundelius, B., et al.: 1997, \journal{Astrophys. J.} \vol{484}, 180 

Tanaka, T., Menou, K., Haiman, Z.: 2012, 
\journal{MNRAS} \vol{420}, 705

Valtonen, M.J., et al.: 2008, \journal{Nature} \vol{452}, 851

Valtonen, M.J., et al.: 2016, \journal{Astrophys. J. Letters} \vol{819}, L37

Valtonen, M.J., et al.: 2019, \journal{Astrophys. J.} \vol{882}, 88

Vaughan, S., et al.: 2016, \journal{MNRAS} \vol{461}, 3145

Volonteri, M.,  Haardt, F., Madau, P.: 2003, \journal{Astrophys. J.} \vol{582}, 559

Wang, J.M., Bon, E.: 2020, \journal{Astron.  Astrophys.} \vol{643}, L9

%% Xu, D., Komossa, S. 2009, \journal{Astrophys. J.}, 705, L20

Yu, Q., Lu, Y.: 2001, \journal{Astron.  Astrophys.} \vol{377}, 17

Zola, S., et al.: 2020, \journal{Astron. Telegram} \vol{13637}, 1 

\endreferences

\end{document}